\pdfoutput=1
\documentclass[12pt,a4paper]{article}

\usepackage{ifthen} 
\newboolean{pdflatex}
\setboolean{pdflatex}{true} 

\newboolean{articletitles}
\setboolean{articletitles}{true} 

\newboolean{uprightparticles}
\setboolean{uprightparticles}{false} 

\def\paperauthors{BESIII and LHCb collaborations} 
\def\paperasciititle{Precise measurement of the CKM angle gamma with a novel approach} 
\def\papertitle{Precise measurement of the CKM angle $\gamma$ with a novel approach} 
\def\paperkeywords{{High Energy Physics}, {BESII}, {LHCb}} 
\def\papercopyright{\the\year\ CERN for the benefit of the LHCb collaboration and \\ IHEP for the benefit of the BESIII collaboration} 
\def\paperlicence{CC BY 4.0 licence}
\def\paperlicenceurl{https://creativecommons.org/licenses/by/4.0/}

\newif\ifEnableSectionTOCLinks
\EnableSectionTOCLinksfalse 

\usepackage[top=1in, bottom=1.25in, left=1in, right=1in]{geometry}

\usepackage{microtype}
\usepackage{lineno}  
\usepackage{xspace} 
\usepackage{caption} 

\usepackage{graphicx}  
\usepackage{color}
\usepackage{colortbl}
\graphicspath{{./figs/}} 

\usepackage{amsmath} 
\usepackage{amssymb}
\usepackage{amsfonts}
\usepackage{upgreek} 

\newcommand*\patchAmsMathEnvironmentForLineno[1]{%
\expandafter\let\csname old#1\expandafter\endcsname\csname #1\endcsname
\expandafter\let\csname oldend#1\expandafter\endcsname\csname
end#1\endcsname
 \renewenvironment{#1}%
   {\linenomath\csname old#1\endcsname}%
   {\csname oldend#1\endcsname\endlinenomath}%
}
\newcommand*\patchBothAmsMathEnvironmentsForLineno[1]{%
  \patchAmsMathEnvironmentForLineno{#1}%
  \patchAmsMathEnvironmentForLineno{#1*}%
}
\AtBeginDocument{%
\patchBothAmsMathEnvironmentsForLineno{equation}%
\patchBothAmsMathEnvironmentsForLineno{align}%
\patchBothAmsMathEnvironmentsForLineno{flalign}%
\patchBothAmsMathEnvironmentsForLineno{alignat}%
\patchBothAmsMathEnvironmentsForLineno{gather}%
\patchBothAmsMathEnvironmentsForLineno{multline}%
\patchBothAmsMathEnvironmentsForLineno{eqnarray}%
}

\usepackage[pdftex,
            pdfauthor={\paperauthors},
            pdftitle={\paperasciititle},
            pdfkeywords={\paperkeywords}]{hyperref}
\usepackage{hyperxmp}
\hypersetup{
    pdfcopyright={Copyright (C) \papercopyright},
    pdflicenseurl={\paperlicenceurl}
}

\usepackage[colorinlistoftodos,textsize=scriptsize]{todonotes}

\usepackage[bottom,flushmargin,hang,multiple]{footmisc}

\usepackage[all]{hypcap} 

\usepackage{xspace}
\usepackage{upgreek}

\def\lhcb   {\mbox{LHCb}\xspace}

\def\besiii {\mbox{BESIII}\xspace}
\def\cleo   {\mbox{CLEO}\xspace}

\def\MagUp {\mbox{\em Mag\kern -0.05em Up}\xspace}

\ifthenelse{\boolean{uprightparticles}}%
{

 \def\Ppi         {\ensuremath{\uppi}\xspace}

 \def\Ppsi        {\ensuremath{\uppsi}\xspace}

 \def\PDelta      {\ensuremath{\Delta}\xspace}
 \def\PXi         {\ensuremath{\Xi}\xspace}
 \def\PLambda     {\ensuremath{\Lambda}\xspace}
 \def\PSigma      {\ensuremath{\Sigma}\xspace}
 \def\POmega      {\ensuremath{\Omega}\xspace}
 \def\PUpsilon    {\ensuremath{\Upsilon}\xspace}
 \let\oldPi\Pi
 \def\PPi         {\ensuremath{\oldPi}\xspace}

 \def\PB      {\ensuremath{\mathrm{B}}\xspace}
 \def\PD      {\ensuremath{\mathrm{D}}\xspace}

 \def\PK      {\ensuremath{\mathrm{K}}\xspace}
 \def\Pb      {\ensuremath{\mathrm{b}}\xspace}
 \def\Pc      {\ensuremath{\mathrm{c}}\xspace}
 \def\Pd      {\ensuremath{\mathrm{d}}\xspace}
 \def\Pe      {\ensuremath{\mathrm{e}}\xspace}
 \def\Ps      {\ensuremath{\mathrm{s}}\xspace}
 
 \def\Pu      {\ensuremath{\mathrm{u}}\xspace}
 \def\thebaroffset{0.0em}
}
{

 \def\Ppi         {\ensuremath{\pi}\xspace}

 \def\Ppsi        {\ensuremath{\psi}\xspace}
 
 \mathchardef\PDelta="7101
 \mathchardef\PXi="7104
 \mathchardef\PLambda="7103
 \mathchardef\PSigma="7106
 \mathchardef\POmega="710A
 \mathchardef\PUpsilon="7107
 \mathchardef\PPi="7105
 \def\PB      {\ensuremath{B}\xspace}
 \def\PD      {\ensuremath{D}\xspace}

 \def\PK      {\ensuremath{K}\xspace}
 \def\Pb      {\ensuremath{b}\xspace}
 \def\Pc      {\ensuremath{c}\xspace}
 \def\Pd      {\ensuremath{d}\xspace}
 \def\Pe      {\ensuremath{e}\xspace}
 \def\Ps      {\ensuremath{s}\xspace}
 
 \def\Pu      {\ensuremath{u}\xspace}
 \def\thebaroffset{0.18em}
}
\newcommand{\offsetoverline}[2][\thebaroffset]{\kern #1\overline{\kern -#1 #2}}%

\makeatletter
\ifcase \@ptsize \relax
  \newcommand{\miniscule}{\@setfontsize\miniscule{4}{5}}
\or
  \newcommand{\miniscule}{\@setfontsize\miniscule{5}{6}}
\or
  \newcommand{\miniscule}{\@setfontsize\miniscule{5}{6}}
\fi
\makeatother

\DeclareRobustCommand{\optbar}[1]{\shortstack{{\miniscule (\rule[.5ex]{1.25em}{.18mm})}
  \\ [-.7ex] $#1$}}

\def\epem       {{\ensuremath{\Pe^+\Pe^-}}\xspace}

\def\uquark    {{\ensuremath{\Pu}}\xspace}
\def\uquarkbar {{\ensuremath{\overline \uquark}}\xspace}

\def\dquark    {{\ensuremath{\Pd}}\xspace}

\def\squark    {{\ensuremath{\Ps}}\xspace}

\def\cquark    {{\ensuremath{\Pc}}\xspace}
\def\cquarkbar {{\ensuremath{\overline \cquark}}\xspace}

\def\bquark    {{\ensuremath{\Pb}}\xspace}

\def\pion   {{\ensuremath{\Ppi}}\xspace}

\def\pip    {{\ensuremath{\pion^+}}\xspace}
\def\pim    {{\ensuremath{\pion^-}}\xspace}
\def\pipm   {{\ensuremath{\pion^\pm}}\xspace}

\def\kaon    {{\ensuremath{\PK}}\xspace}

\def\KorKbar {\kern \thebaroffset\optbar{\kern -\thebaroffset \PK}{}\xspace}

\def\Kp      {{\ensuremath{\kaon^+}}\xspace}
\def\Km      {{\ensuremath{\kaon^-}}\xspace}
\def\Kpm     {{\ensuremath{\kaon^\pm}}\xspace}

\def\KS      {{\ensuremath{\kaon^0_{\mathrm{S}}}}\xspace}

\def\KL      {{\ensuremath{\kaon^0_{\mathrm{L}}}}\xspace}

\def\Dbar    {{\ensuremath{\offsetoverline{\PD}}}\xspace}
\def\D       {{\ensuremath{\PD}}\xspace}
\def\Db      {{\ensuremath{\Dbar}}\xspace}
\def\DorDbar {\kern \thebaroffset\optbar{\kern -\thebaroffset \PD}\xspace}
\def\Dz      {{\ensuremath{\D^0}}\xspace}
\def\Dzb     {{\ensuremath{\Dbar{}^0}}\xspace}
\def\Dp      {{\ensuremath{\D^+}}\xspace}
\def\Dm      {{\ensuremath{\D^-}}\xspace}

\def\DpDm    {\ensuremath{\Dp {\kern -0.16em \Dm}}\xspace}

\def\B       {{\ensuremath{\PB}}\xspace}

\def\BorBbar {\kern \thebaroffset\optbar{\kern -\thebaroffset \PB}\xspace}

\def\Bd      {{\ensuremath{\B^0}}\xspace}

\def\BdorBdbar {\kern \thebaroffset\optbar{\kern -\thebaroffset \Bd}\xspace}
\def\Bu      {{\ensuremath{\B^+}}\xspace}
\def\Bub     {{\ensuremath{\B^-}}\xspace}
\def\Bp      {{\ensuremath{\Bu}}\xspace}
\def\Bm      {{\ensuremath{\Bub}}\xspace}
\def\Bpm     {{\ensuremath{\B^\pm}}\xspace}

\def\Bs      {{\ensuremath{\B^0_\squark}}\xspace}

\def\BsorBsbar {\kern \thebaroffset\optbar{\kern -\thebaroffset \Bs}\xspace}

\def\psiprpr  {{\ensuremath{\Ppsi(3770)}}\xspace}

\def\Y#1S{\ensuremath{\PUpsilon{(#1S)}}\xspace}

\def\LorLbar     {\kern \thebaroffset\optbar{\kern -\thebaroffset \PLambda}\xspace}

\newcommand{\decay}[2]{\mbox{\ensuremath{#1\!\to #2}}\xspace}

\def\to                 {\ensuremath{\rightarrow}\xspace}

\def\CP                {{\ensuremath{C\!P}}\xspace}

\def\Vud  {{\ensuremath{V_{\uquark\dquark}^{\phantom{\ast}}}}\xspace}
\def\Vcd  {{\ensuremath{V_{\cquark\dquark}^{\phantom{\ast}}}}\xspace}

\def\Vubs  {{\ensuremath{V_{\uquark\bquark}^\ast}}\xspace}
\def\Vcbs  {{\ensuremath{V_{\cquark\bquark}^\ast}}\xspace}

\def\AT#1     {\ensuremath{A_{\mathrm{T}}^{#1}}\xspace}           

\def\C#1      {\ensuremath{\mathcal{C}_{#1}}\xspace}                       
\def\Cp#1     {\ensuremath{\mathcal{C}_{#1}^{'}}\xspace}                    
\def\Ceff#1   {\ensuremath{\mathcal{C}_{#1}^{\mathrm{(eff)}}}\xspace}        
\def\Cpeff#1  {\ensuremath{\mathcal{C}_{#1}^{'\mathrm{(eff)}}}\xspace}       
\def\Ope#1    {\ensuremath{\mathcal{O}_{#1}}\xspace}                       
\def\Opep#1   {\ensuremath{\mathcal{O}_{#1}^{'}}\xspace}                    

\newcommand{\aunit}[1]{\ensuremath{\text{\,#1}}}

\newcommand{\tev}{\aunit{Te\kern -0.1em V}\xspace}
\newcommand{\gev}{\aunit{Ge\kern -0.1em V}\xspace}
\newcommand{\mev}{\aunit{Me\kern -0.1em V}\xspace}
\newcommand{\kev}{\aunit{ke\kern -0.1em V}\xspace}
\newcommand{\ev}{\aunit{e\kern -0.1em V}\xspace}

\newcommand{\mevc}{\ensuremath{\aunit{Me\kern -0.1em V\!/}c}\xspace}
\newcommand{\gevc}{\ensuremath{\aunit{Ge\kern -0.1em V\!/}c}\xspace}
\newcommand{\mevcc}{\ensuremath{\aunit{Me\kern -0.1em V\!/}c^2}\xspace}
\newcommand{\gevcc}{\ensuremath{\aunit{Ge\kern -0.1em V\!/}c^2}\xspace}

\def\fb   {\ensuremath{\aunit{fb}}\xspace}
\def\invfb   {\ensuremath{\fb^{-1}}\xspace}

\newcommand{\chisqndf}{\ensuremath{\chi^2/\mathrm{ndf}}\xspace}

\def\deriv {\ensuremath{\mathrm{d}}}

\def\gsim{{~\raise.15em\hbox{$>$}\kern-.85em
          \lower.35em\hbox{$\sim$}~}\xspace}
\def\lsim{{~\raise.15em\hbox{$<$}\kern-.85em
          \lower.35em\hbox{$\sim$}~}\xspace}

\def\sPlot{\mbox{\em sPlot}\xspace}

\def\sqs   {\ensuremath{\protect\sqrt{s}}\xspace}

\def\degrees{\ensuremath{^{\circ}}\xspace}

\def\evtgen     {\mbox{\textsc{EvtGen}}\xspace}

\def\geant      {\mbox{\textsc{Geant4}}\xspace}

\def\photos     {\mbox{\textsc{Photos}}\xspace}

\def\pythia     {\mbox{\textsc{Pythia}}\xspace}

\def\tell1  {TELL1\xspace}
\def\ukl1   {UKL1\xspace}

\newcommand{\eg}{\mbox{\itshape e.g.}\xspace}

\newcommand{\vs}{\mbox{\itshape vs.}\xspace}

\newcommand{\lhcborcid}[1]{\href{https://orcid.org/#1}{\hspace*{0.1em}\raisebox{-0.45ex}{\includegraphics[width=1em]{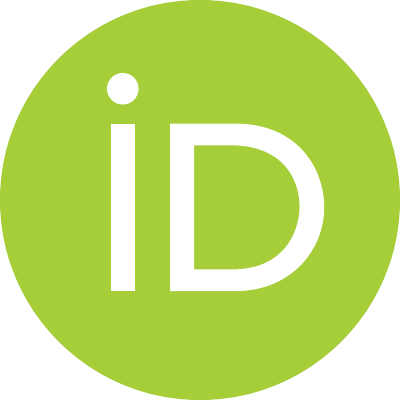}}}}

\def\rb       {\ensuremath{r_{B}}\xspace}

\def\db       {\ensuremath{\delta_{B}}\xspace}

\def\rdk      {\ensuremath{r_{B}^{DK}}\xspace}
\def\rdpi     {\ensuremath{r_{B}^{D\pi}}\xspace}

\def\ddk      {\ensuremath{\delta_{B}^{DK}}\xspace}
\def\ddpi     {\ensuremath{\delta_{B}^{D\pi}}\xspace}

\def\pd       {\ensuremath{p_{D}}\xspace}
\def\pdb      {\ensuremath{\overline{p}_{D}}\xspace}

\def\xpdk     {\ensuremath{x_{+}^{DK}}\xspace}
\def\xmdk     {\ensuremath{x_{-}^{DK}}\xspace}
\def\ypdk     {\ensuremath{y_{+}^{DK}}\xspace}
\def\ymdk     {\ensuremath{y_{-}^{DK}}\xspace}

\def\xpmdk    {\ensuremath{x_{\pm}^{DK}}\xspace}
\def\ypmdk    {\ensuremath{y_{\pm}^{DK}}\xspace}
\def\xpmdh    {\ensuremath{x_{\pm}^{Dh}}\xspace}
\def\ypmdh    {\ensuremath{y_{\pm}^{Dh}}\xspace}

\def\xxi      {\ensuremath{x_{\xi}^{D\pi}}\xspace}
\def\yxi      {\ensuremath{y_{\xi}^{D\pi}}\xspace}

\def\Cph      {\ensuremath{\mathcal{C}}\xspace}
\def\Sph      {\ensuremath{\mathcal{S}}\xspace}

\def\hp       {\ensuremath{h^{+}}}
\def\hm       {\ensuremath{h^{-}}}

\def\KSL      {\ensuremath{K^{0}_{\mathrm{S/L}}}}

\def\KShh     {\ensuremath{\KS\hp\hm}}

\def\Nobs{\ensuremath{\mathcal{N}}}
\def\Nobsbar{\ensuremath{\smash{\overline{\mathcal{N}}}}} 

\hypersetup{
  colorlinks   = true, 
  urlcolor     = blue, 
  linkcolor    = blue, 
  citecolor    = red   
}

\ifEnableSectionTOCLinks
    \usepackage[explicit]{titlesec} 
    
    \let\oldcontentsline\contentsline
    \renewcommand

    \titleformat{\section}{\normalfont\Large\bf}{\hyperlink{tocsection.\thesection}{{\thesection} \parbox[t]{\dimexpr\textwidth-1pc}{#1}}}{1pc}{}

    \titleformat{\subsection}{\normalfont\bf}{\hyperlink{tocsubsection.\thesubsection}{{\thesubsection} \parbox[t]{\dimexpr\textwidth-1pc}{#1}}}{1pc}{}

    \titleformat{name=\section,numberless}[display]{}{}{0pt}{\normalfont\Huge\bfseries #1}
\fi

\usepackage{cite} 
\usepackage{LHCb/mciteplus}

\newcommand{\BESIIIorcid}[1]{\href{https://orcid.org/#1}{\hspace*{0.1em}\raisebox{-0.45ex}{\includegraphics[width=1em]{LHCb/Figures/orcidIcon.pdf}}}}

\begin{document}

\renewcommand{\thefootnote}{\fnsymbol{footnote}}
\setcounter{footnote}{1}


\begin{titlepage}
\pagenumbering{roman}

\vspace*{-1.5cm}
\centerline{\large EUROPEAN ORGANIZATION FOR NUCLEAR RESEARCH (CERN)}
\centerline{\large INSTITUTE OF HIGH ENERGY PHYSICS (IHEP)}
\vspace*{1.5cm}
\noindent
\begin{tabular*}{\linewidth}{lc@{\extracolsep{\fill}}r@{\extracolsep{0pt}}}
\ifthenelse{\boolean{pdflatex}}
{\vspace*{-1.5cm}\mbox{\!\!\!
 \includegraphics[width=.09\textwidth,angle=90]{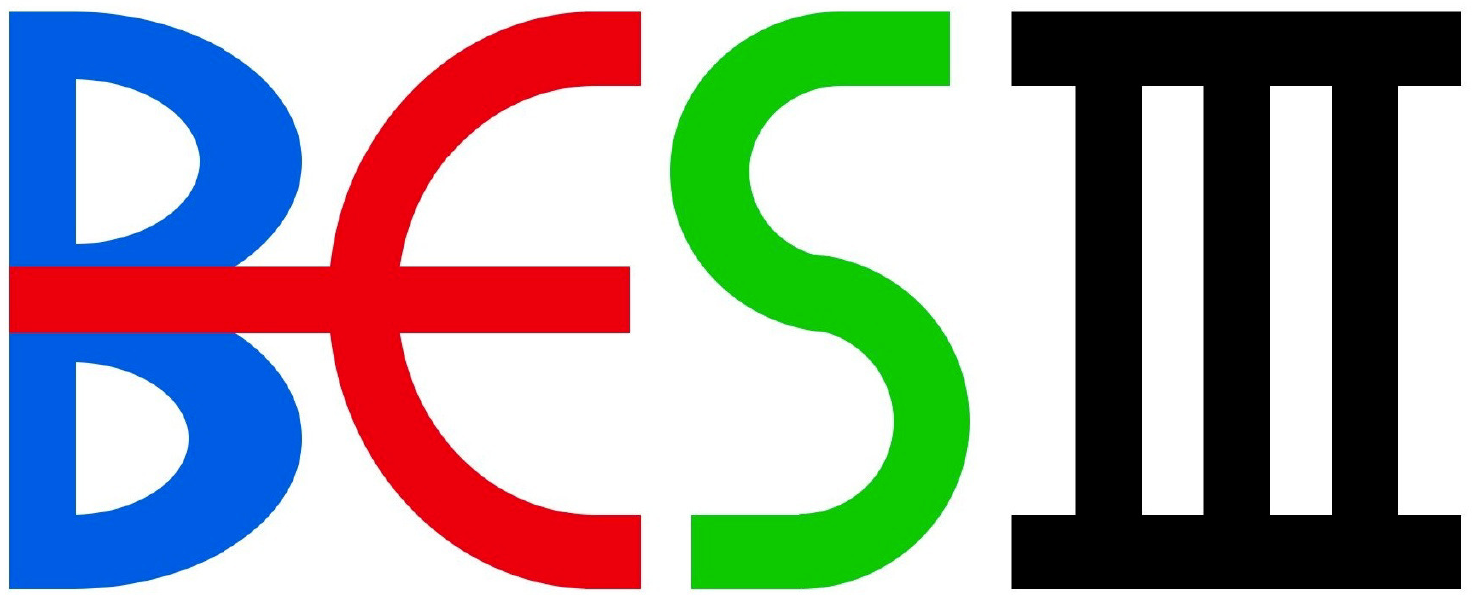}~
\includegraphics[width=.14\textwidth]{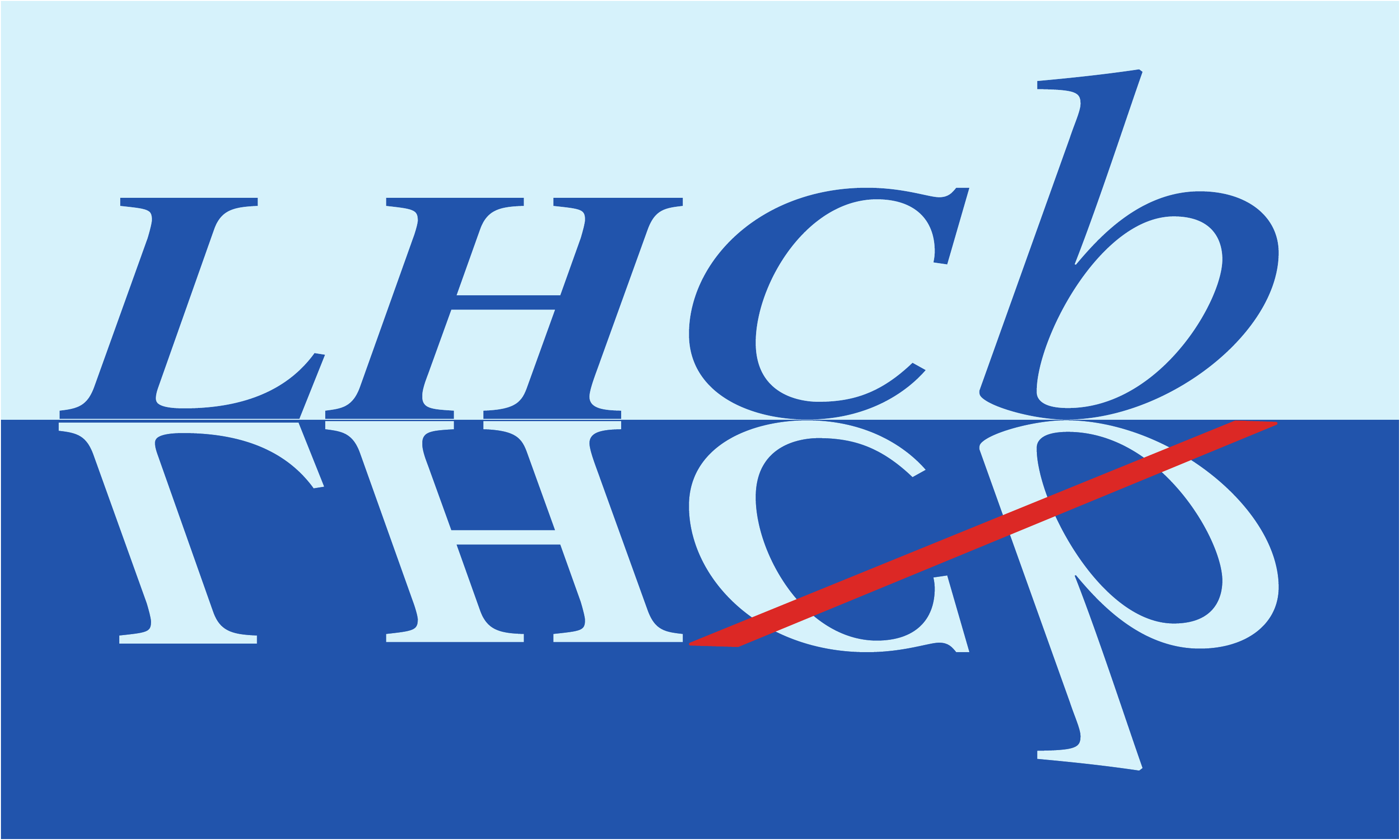}} & &}%
{\vspace*{-1.2cm}\mbox{\!\!\!
 \includegraphics[width=.08\textwidth,angle=90]{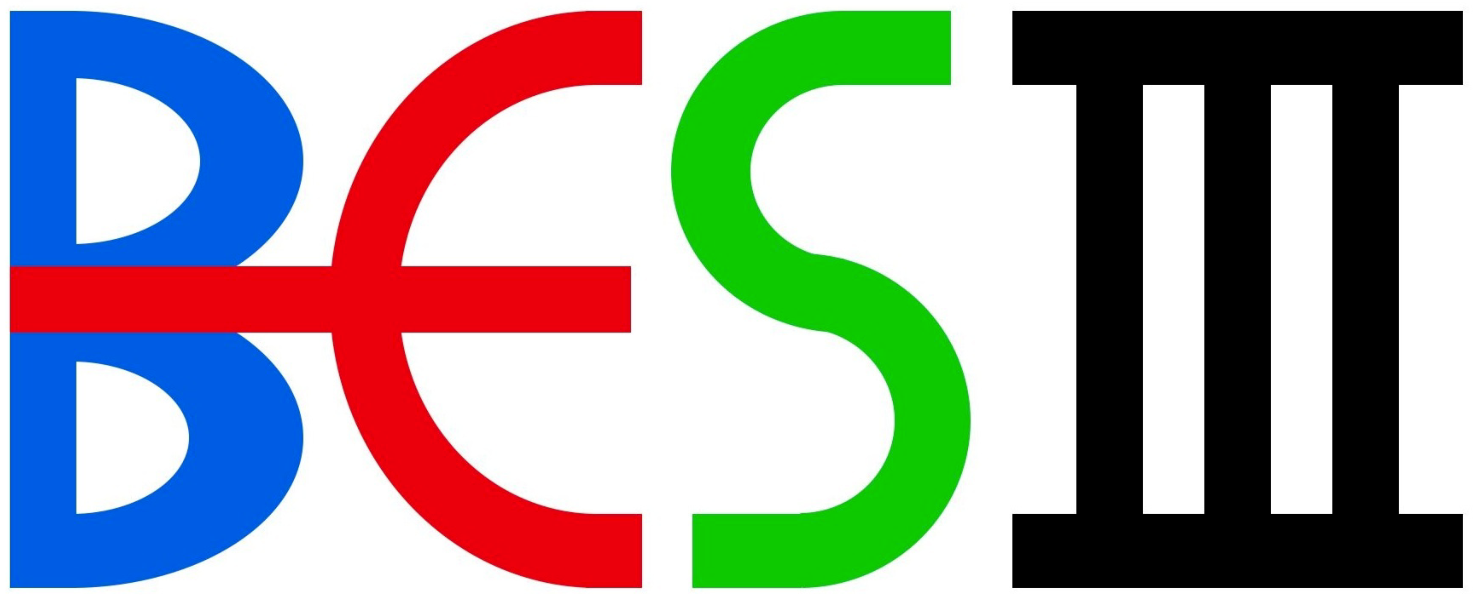}~
\includegraphics[width=.12\textwidth]{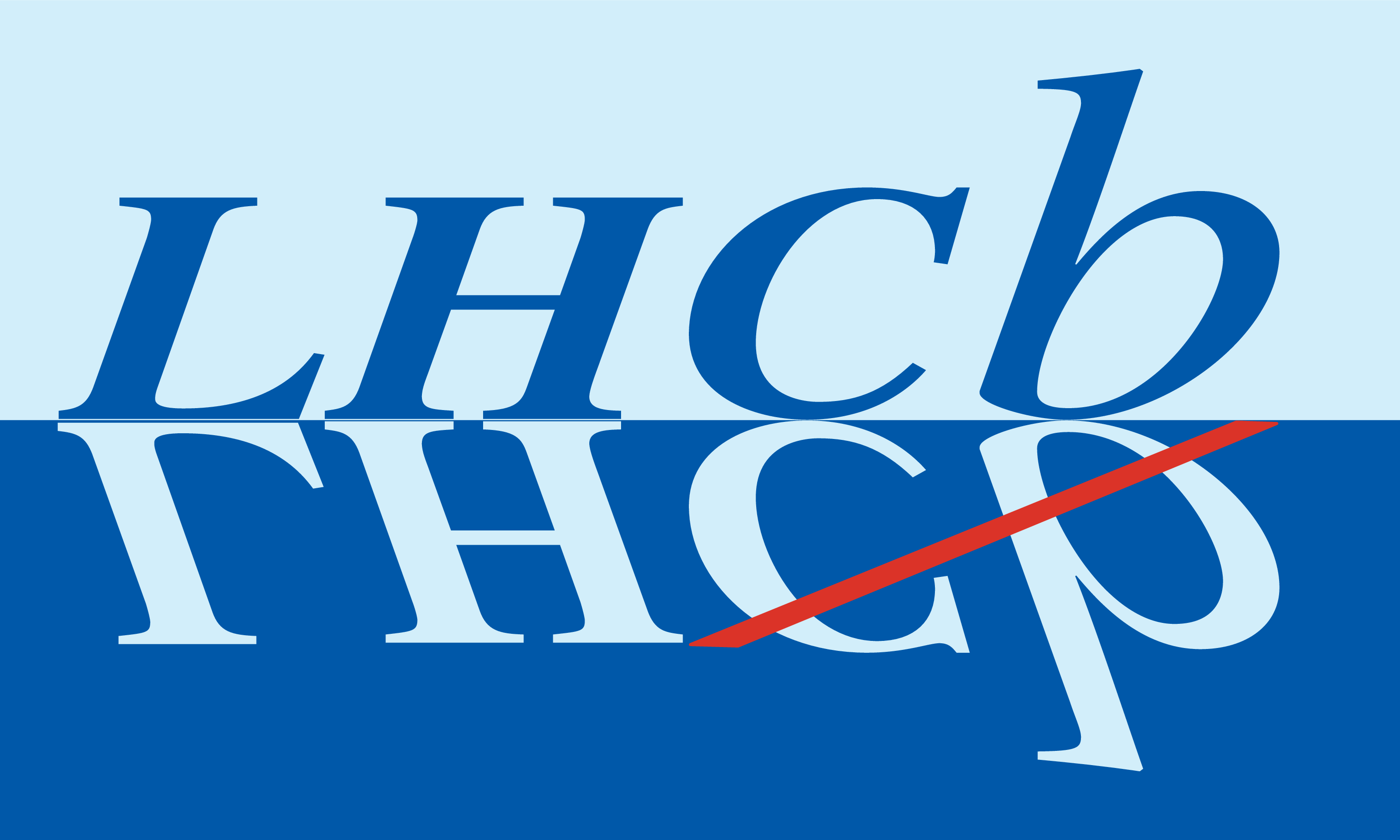}} & &}%
\\
 & & CERN-EP-2026-068 \\  
 & & LHCb-PAPER-2025-064 \\  
 & & April 7, 2026 \\
\end{tabular*}

\vspace*{2.0cm}

{\normalfont\bfseries\boldmath\huge
\begin{center}
  \papertitle 
\end{center}
}

\vspace*{1.5cm}

\begin{center}
\paperauthors\footnote{Authors are listed at the end of this Letter.}
\end{center}

\vspace{\fill}

\begin{abstract}
  \noindent
  A measurement of the CKM angle $\gamma$ is performed by applying a novel, unbinned, model-independent approach to datasets of electron-positron collisions collected by the \besiii experiment and proton-proton collisions by the \lhcb experiment,
  corresponding to integrated luminosities of 8\invfb and 9\invfb, respectively.
  The \CP-violating phase $\gamma$ is determined from $\decay{\Bpm}{D(\to\KS h^{\prime+}h^{\prime-}) h^{\pm}}$ decays in \lhcb data, where $h^{(\prime)}$ is either a pion or kaon,
  while the corresponding strong-phase parameters are measured using doubly tagged ${D\to K_{\rm S/L}^0 h^{\prime+} h^{\prime-}}$ decays in the quantum-correlated $D\Db$ system present in \besiii data.
  A joint fit to both datasets, which allows for a simultaneous determination of the associated \CP-violating observables and strong-phase parameters, yields ${\gamma = (71.3\pm 5.0)\degrees}$.
  The result is the most precise to date and consistent with previous measurements and world averages.
\end{abstract}

\vspace*{2.0cm}

\begin{center}
  Submitted to Phys.~Rev.~Lett.
\end{center}

\vspace{\fill}

{\footnotesize 
\begin{center}
\copyright~\papercopyright. \href{\paperlicenceurl}{\paperlicence}.
\end{center}
}
\vspace*{2mm}

\end{titlepage}


\newpage
\setcounter{page}{2}
\mbox{~}


\renewcommand{\thefootnote}{\arabic{footnote}}
\setcounter{footnote}{0}


\cleardoublepage


\pagestyle{plain} 
\setcounter{page}{1}
\pagenumbering{arabic}


Violation of symmetry under charge-parity (\CP) transformation 
is a necessary ingredient to explain the observed matter-antimatter asymmetry of the Universe~\cite{Sakharov:1967dj}. In the Standard Model of particle physics~(SM), \CP violation is explained by a complex phase in the Cabibbo--Kobayashi--Maskawa (CKM) matrix~\cite{Cabibbo:1963yz,*Kobayashi:1973fv}, which governs quark flavor mixing in the weak interaction. 
The unitarity of this matrix leads to relationships that can be visualized as triangles in the complex plane, known as Unitarity Triangles.
Among their internal angles, ${\gamma=\phi_3\equiv\arg(-\Vud \Vubs/\Vcd \Vcbs)}$, with $V_{xy}$ representing the transition amplitude from quark $x$ to quark $y$, is of particular interest.
It can be determined directly either through the interference between $\decay{b}{c\uquarkbar s}$ and $\decay{b}{ u\cquarkbar s}$ transitions at leading order (tree-level), \eg $\decay{\Bpm}{D\Kpm}$ decays, with negligible theoretical uncertainties~\cite{Brod:2013sga,Brod:2014qwa}, or in charmless $B$-meson decays~\cite{Fleischer:1999pa,Fleischer:2007hj,Fleischer:2010ib,Fleischer:2022rkm,Rey-LeLorier:2011ltd,Bhattacharya:2014eca,Bhattacharya:2023pef}. 
In contrast, indirect measurements of $\gamma$ often involve loop-mediated processes, which are potentially sensitive to physics beyond the SM.
Comparisons between direct tree-level measurements and indirect measurements provide a powerful test of CKM unitarity and could reveal new sources of \CP violation. 
Currently, the precision of the direct measurement~\cite{LHCb-CONF-2025-003} is about a factor of $3$ worse than those  using indirect constraints~\cite{CKMfitter2005,CKMfitter2015,UTfit-UT} assuming CKM unitarity. 
Consequently, improving the precision of the direct measurements of $\gamma$ is a crucial goal in particle physics.

Direct measurements of $\gamma$ are primarily obtained from $\decay{B}{D^{(*)}K^{(*)}}$ decays, 
where the global precision is currently dominated by the \lhcb experiment, which has recently determined ${\gamma=(62.8\pm2.6){\degrees}}$~\cite{LHCb-CONF-2025-003}.
The most precise single measurement to date comes from the analysis of $\decay{\Bpm}{Dh^{\pm}}$ decays, and $\decay{D}{\KS h^{\prime +}h^{\prime -}}$ decay with $h^{(\prime)}$ being a kaon or pion, where $D$ denotes a coherent superposition of $\Dz$ and $\Dzb$ mesons.
The decay phase-space amplitudes can be described in terms of the Dalitz-plot~\cite{Dalitz:1953cp,Fabri:1954zz} variables,
defined here as the $\KS h^{\prime+}$ and $\KS h^{\prime-}$ squared masses.
The measurement exploits strong-phase variation across the Dalitz plot using a binned phase-space approach~\cite{BondarProc,Giri:2003ty,Bondar:2008hh}, yielding ${\gamma=(68.7^{+5.2}_{-5.1}){\degrees}}$~\cite{LHCb-PAPER-2020-019}. 
To determine $\gamma$ from $\decay{\Bpm}{Dh^{\pm}}$ data,
information of the strong-phase difference between $\Dz$ and $\Dzb$ decays to ${\KS h^{\prime+}h^{\prime-}}$ is required,
which has been measured independently in quantum-correlated $D\Db$ decays produced in $\epem$ collisions by the \cleo~\cite{CLEO:2010iul} and \besiii experiments~\cite{BESIII:2020hlg,BESIII:2020khq,BESIII:2020hpo}.
The binned technique provides an unbiased determination of $\gamma$ by avoiding model-dependent assumptions. 
However, it sacrifices some statistical sensitivity because events within each bin are averaged for strong-phase values and decay densities, thereby neglecting intra-bin information.
Current binning schemes are estimated to retain approximately 85\% of the sensitivity to $\gamma$~\cite{CLEO:2010iul}.
Advanced analysis techniques that increase sensitivity while preserving model independence are therefore essential for precision tests of CKM unitarity. 

The decay rate of $\decay{\Bpm}{Dh^{\pm}}$, when ignoring $D$ mixing and \CP violation in $D$ decays, is given by 
\begin{equation}
\begin{aligned}
    p_{{B^{-}}}(\mathbf{z}) & \propto p_{D}(\mathbf{z}) + \left[(x_{-}^{Dh})^2 + (y_{-}^{Dh})^2\right] p_{D}(\mathbf{\bar{z}})+ 2\left[x_{-}^{Dh}\Cph(\mathbf{z}) + y_{-}^{Dh}\Sph(\mathbf{z})\right], \label{eq:Bmrate}  \\ 
    p_{{B^{+}}}(\mathbf{z}) & \propto p_{D}(\mathbf{\bar{z}}) + \left[(x_{+}^{Dh})^2 + (y_{+}^{Dh})^2\right]p_{D}(\mathbf{z})+ 2\left[x_{+}^{Dh}\Cph(\mathbf{z}) - y_{+}^{Dh}\Sph(\mathbf{z})\right], 
\end{aligned}
\end{equation}
where $p_{\Bpm}$ and $\pd$ are the decay rates of $\decay{\Bpm}{Dh^{\pm}}$ and $\decay{\Dz}{\KShh}$, respectively. The coordinates ${\mathbf{z}\equiv(m_{\KS h^{\prime+}}^2,~m_{\KS h^{\prime-}}^2)}$ represents a point on the Dalitz plane of the $\decay{\Dz}{\KS h^{\prime+}h^{\prime-}}$ decay and ${\mathbf{\bar{z}}\equiv (m_{\KS h^{\prime-}}^2,~m_{\KS h^{\prime+}}^2)}$ denotes the charge-conjugated point.
The \CP-violating parameters of \Bpm, $\xpmdh$ and $\ypmdh$, are defined as ${\xpmdh + i\ypmdh \equiv \rb^{Dh} e^{i(\db^{Dh}\pm\gamma)}}$, where $\rb^{Dh}$ and $\db^{Dh}$ denote the amplitude ratio and strong-phase difference between the suppressed decay $\decay{\Bm}{\Dzb h^-}$ and the favored decay $\decay{\Bm}{\Dz h^-}$. 
As the $\decay{\Bpm}{D\Kpm}$ and $\decay{\Bpm}{D\pipm}$ decays share the same weak phase $\gamma$, the decays $\decay{\Bpm}{D\pipm}$ can be described by two additional parameters, ${\xxi + i\yxi\equiv\left(\rdpi\middle/\rdk\right) \cdot e^{i(\ddpi - \ddk)}}$, besides the four $\decay{\Bpm}{D\Kpm}$ parameters, $\xpmdk$, $\ypmdk$.

The magnitude of the ratio between the amplitudes of $\Dzb$ and $\Dz$ decaying to ${\KS h^{\prime+}h^{\prime-}}$ is denoted as ${r_D(\mathbf{z})}$, with ${[r_D(\mathbf{z})]^2 = p_D(\mathbf{\bar{z}})/p_D(\mathbf{z})}$.
The functions $\Cph(\mathbf{z})$ and $\Sph(\mathbf{z})$ are defined as ${\sqrt{\pd(\mathbf{z})\pdb(\overline{\mathbf{z}})}\cos\phi(\mathbf{z})}$ and ${\sqrt{\pd(\mathbf{z})\pdb(\overline{\mathbf{z}})}\sin\phi(\mathbf{z})}$, respectively, with $\mathbf{\phi}(\mathbf{z})$ the strong-phase difference between the two $D$ decays. A key feature of this formalism is that Eq.~\ref{eq:Bmrate} remains valid under integration over the Dalitz plane with an arbitrary weight function $w(\mathbf{z})$. This property forms the basis for an unbiased, model-independent extraction of $\gamma$~\cite{Poluektov:2017zxp}. So the question of how to achieve the best sensitivity to $\gamma$ then translates into finding the optimal set of weight functions. The binned approach is a special case where $w(\mathbf{z})$ is a step function: $w_i(\mathbf{z})=1$ if $\mathbf{z}$ lies within bin $i$, and $0$ otherwise. 
The model-independent integrated strong-phase parameters, ${\{C,~S\}=\int w(\mathbf{z})\{\Cph(\mathbf{z}),\Sph(\mathbf{z})\}\deriv\mathbf{z}}$, can be measured using quantum-correlated $D\Db$ pairs produced in $e^+e^-$ annihilation. Taking both of the $D\Db$ mesons decaying to $\KS h^{\prime+}h^{\prime-}$ as an example, their joint decay rate can be related to the strong-phase parameters as
\begin{equation}
\begin{aligned}
    p_{D\Db}(\mathbf{z}_{D},~\mathbf{z}_{\Dbar}) & \propto p_{D}(\mathbf{z}_{D})p_{D}(\mathbf{\overline{z}}_{\Dbar}) + p_{D}(\mathbf{z}_{\Dbar})p_{D}(\mathbf{\overline{z}}_{D}) - 2\left[\Cph(\mathbf{z}_{D})\Cph(\mathbf{z}_{\Dbar})+\Sph(\mathbf{z}_{D})\Sph(\mathbf{z}_{\Dbar})\right]. \label{eq:Drate} 
\end{aligned}
\end{equation}

\begin{figure}[tb]
    \centering
    \includegraphics[width=0.43\textwidth]{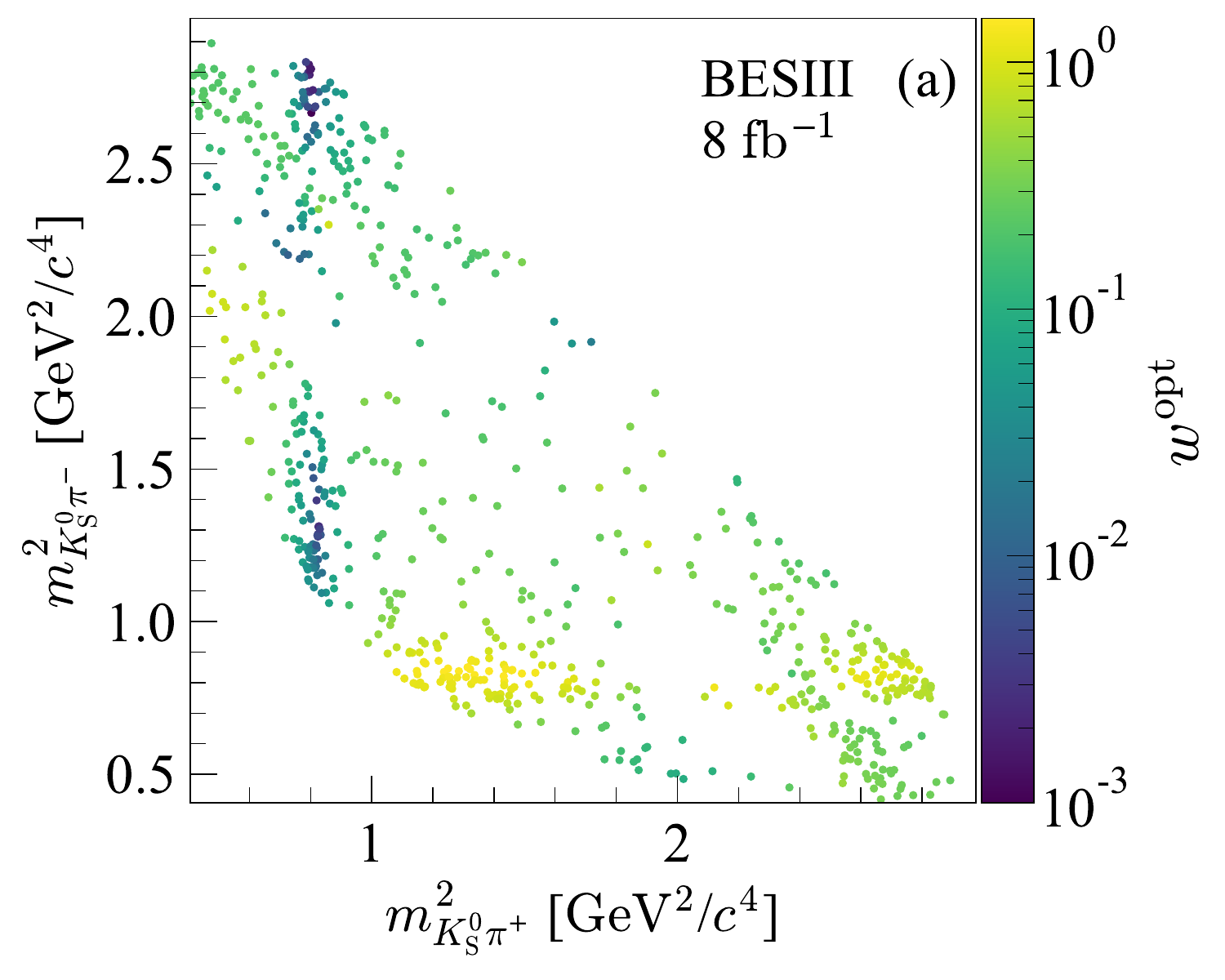}
    \includegraphics[width=0.43\textwidth]{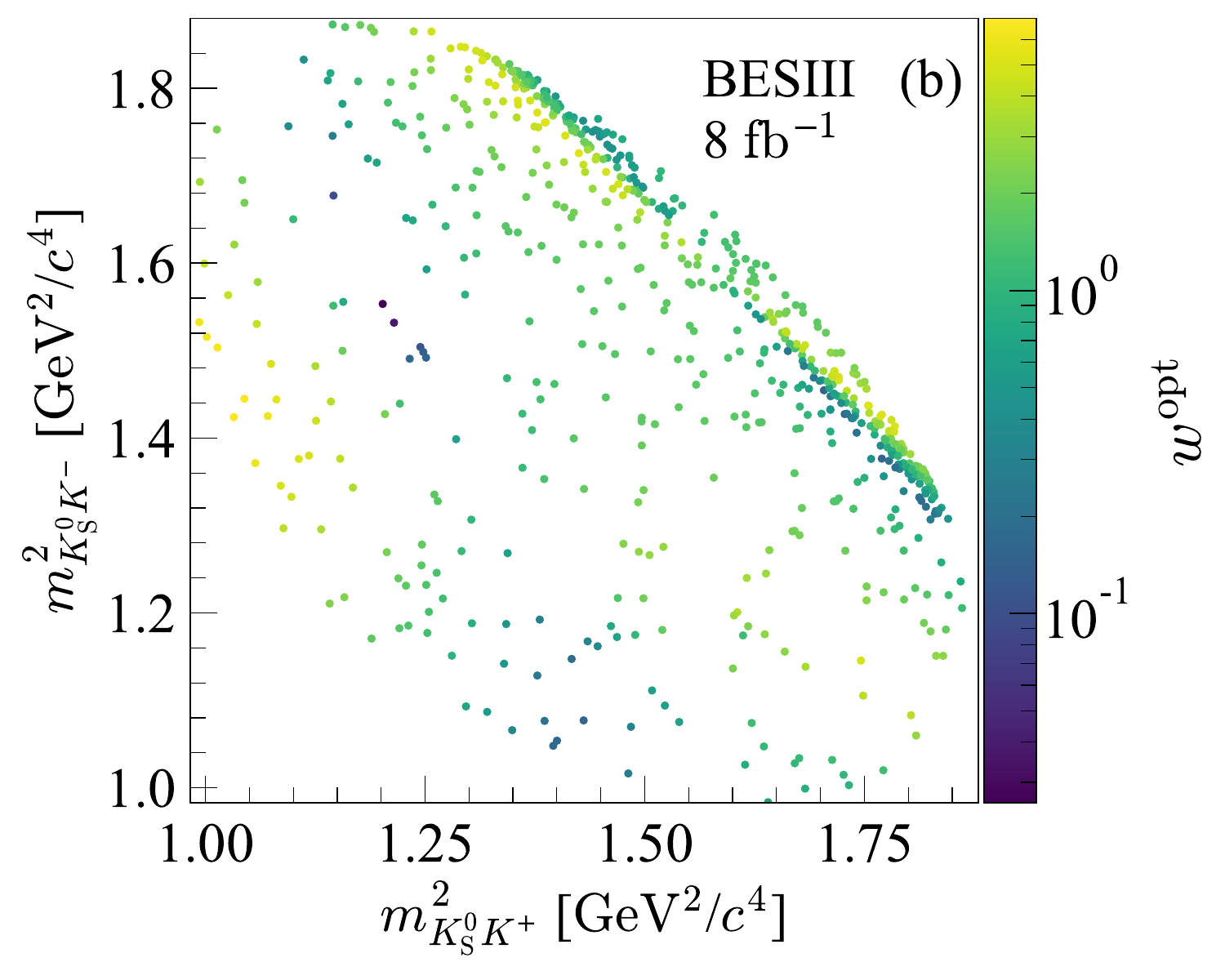}

    \includegraphics[width=0.43\textwidth]{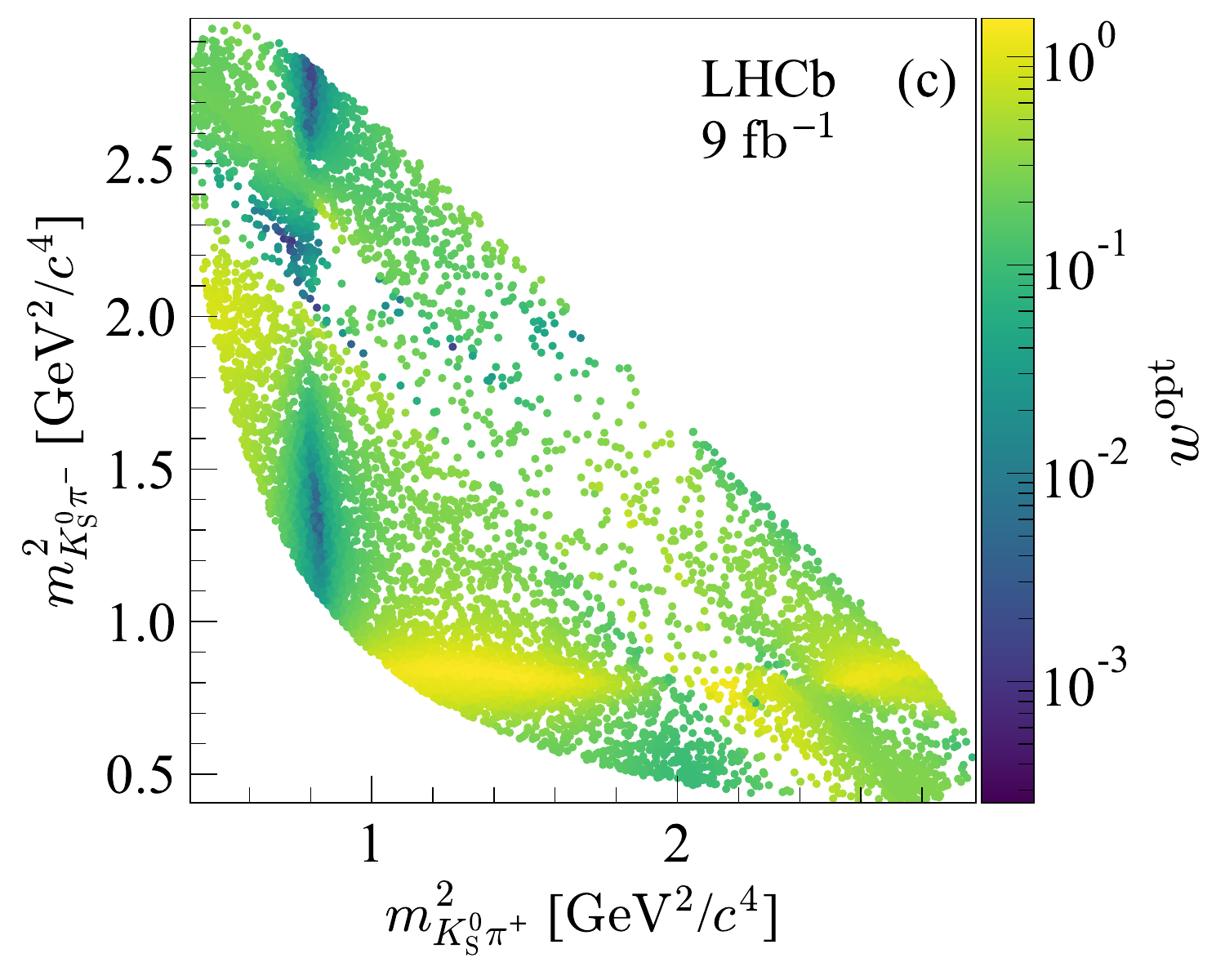}
    \includegraphics[width=0.43\textwidth]{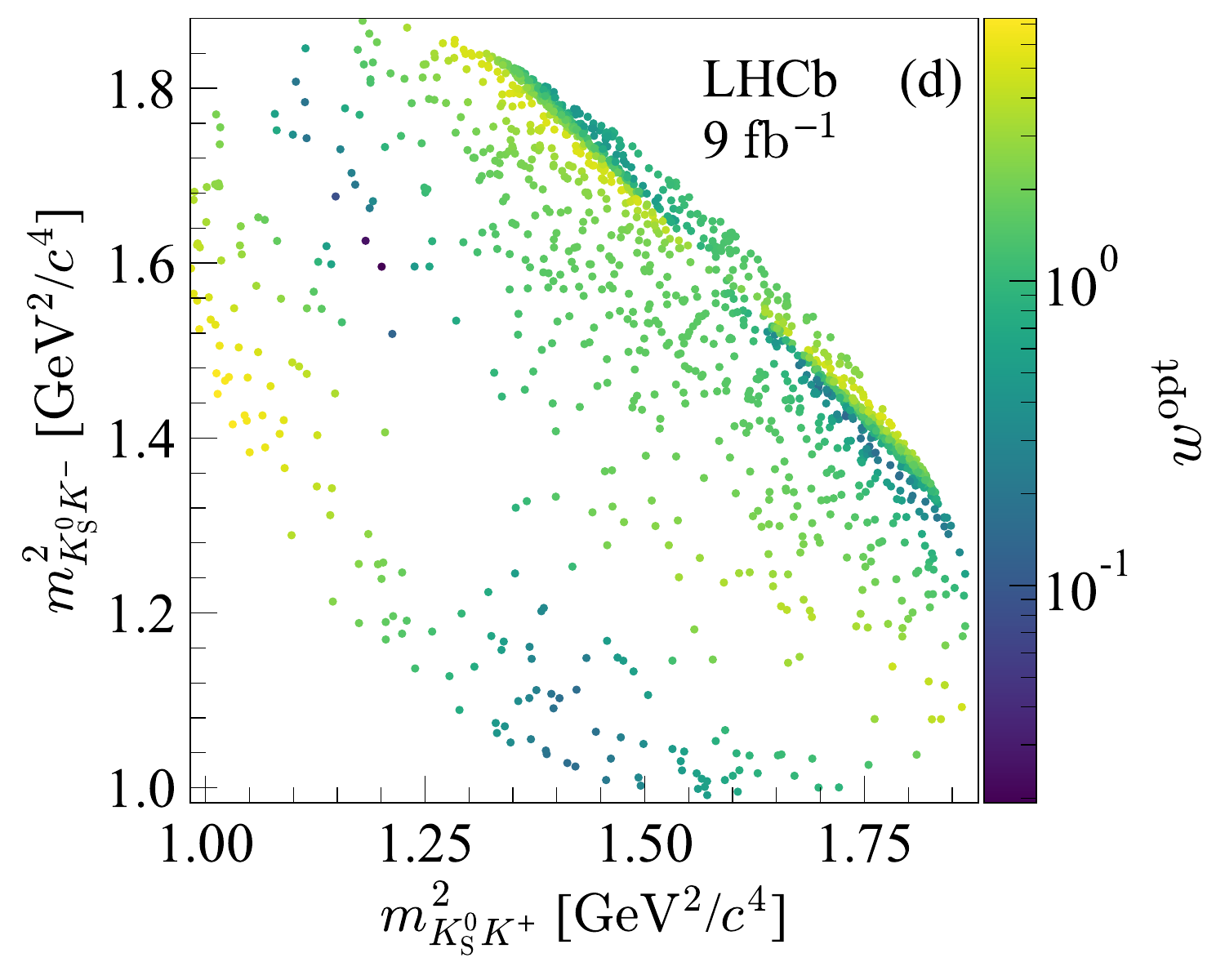}
    \caption{(a) and (b) show the optimal weight $w^{\rm opt}$ for signal candidates from the \besiii data sample with one $D$ decaying to ${\KS\pip\pim}$ and the other to ${\KS\Kp\Km}$, where $\decay{D}{\KS\pip\pim}$ and $\decay{D}{\KS\Kp\Km}$ are the tag mode and signal decay, respectively; (c) and (d) display $w^{\rm opt}$ for signal candidates from the \lhcb data sample for $\decay{B^\pm}{DK^\pm}$ decays with the $\decay{D}{\KS\pip\pim}$ and $\decay{D}{\KS\Kp\Km}$ final states, respectively. 
    }
    \label{fig:opt_weights_data}
\end{figure}

In this Letter, a new measurement of $\gamma$ using a novel set of weight functions based on data from the \besiii and \lhcb collaborations is presented. 
An extensive description of the experimental procedure along with a comprehensive collection of results can be found in a companion article~\cite{LHCb-PAPER-2025-063}.
The method employs an unbinned approach that combines two components to maximize sensitivity. 
The first component, following the concept introduced in Ref.~\cite{Poluektov:2017zxp},
utilizes a Fourier expansion of the strong-phase difference $\mathbf{\phi}(\mathbf{z})$, to capture its variation across the Dalitz plot,
and is calculated from amplitude models of $\decay{D}{\KS h^{\prime+} h^{\prime-}}$ decays~\cite{BaBar:2010nhz,BaBar:2018agf,BaBar:2018cka}. The second component introduced here, $w^{\rm opt}(\mathbf{z})$, is the optimal weight incorporating sensitivity changes due to $r_D(\mathbf{z})$ and experimental effects such as efficiency and background distributions over the Dalitz plot.
The values of $w^{\rm opt}$ for the datasets, shown in Fig.~\ref{fig:opt_weights_data}, enhance statistical sensitivity by assigning greater importance to regions that exhibit larger \CP-violating effects.
While amplitude models are used to construct the weight functions, the $\gamma$ measurement remains model independent as the models only define integration weights without directly parameterizing the data. 
Improved amplitude models may lead to sensitivity gains through more accurate weights, and this approach could therefore benefit from future amplitude studies with larger data samples. 

For a Fourier expansion truncating at the order of $M$, a total of $2\times(2M+1)$ weight functions are defined as
\begin{align}
    \optbar{w}_{\hspace*{-0.2em}n}(\mathbf{z})\equiv \left\{
    \begin{array}{ll}
    w^{\rm opt}(\optbar{\mathbf{z}})\cos\left[k\phi(\mathbf{z})\right], & n = 2k, \\
    w^{\rm opt}(\optbar{\mathbf{z}})\sin\left[k\phi(\mathbf{z})\right], & n = 2k-1,
    \end{array}
    \right.\label{eq:weightfunc}
\end{align}
where $k$ enumerates from 0 (1) to $M$ for the cosine (sine) weights. 
Use of the above set of weight functions is referred to the \emph{optimal Fourier} method throughout this Letter. It combines advantages of both binned and model-dependent approaches: model independence is maintained, while the statistical power of a model-dependent analysis is better exploited by utilizing the detailed variation of both the magnitude and phase of the decay amplitude across phase space.

Experimental observables for \CP-violating effects in $\decay{B^\pm}{DK^\pm}$ decays, denoted by $\Nobs_{n}^{\pm}$ and $\Nobsbar_{\!n}^{\pm}$, are determined by summing signal events over the $D$-decay Dalitz plane with the weight functions,
\begin{align}
\left\{\smash{\optbar{\Nobs}}_{\!n}^{+},~\smash{\optbar{\Nobs}}_{\!n}^{-}\right\}\propto\left\{\sum_{i\in\Bp\to D\Kp}\optbar{w}_{\hspace*{-0.2em}n}(\mathbf{z}_i),~\sum_{i\in\Bm\to D\Km}\optbar{w}_{\hspace*{-0.2em}n}(\mathbf{z}_i)\right\},\label{eq:Bobservables}
\end{align}
where $i$ sums over the corresponding signal candidates in the dataset. With maximum Fourier order $M$ for the $\decay{D}{\KS h^{\prime+}h^{\prime-}}$ channel, there are a total of ${4\times(2M+1)}$ $B$-decay observables.
Among them, $\{C_n,~S_n\}$ are constrained using the double-tag method~\cite{PhysRevLett.56.2140} in the $D\Dbar$ system, where one $D$ meson is reconstructed in the signal decay and the other in three categories of tag mode, including (quasi-)flavor-specific decays, \CP-eigenstate decays,
and self-conjugate decays that are invariant under \CP transformation but are not \CP eigenstates. 
The $D$-decay observables are summarized in Table~\ref{tab:observables_BESIII} by summing over the weight functions corresponding to each signal candidate $i$.
The $B$- and $D$-decay observables can be expressed with the \CP-violating parameters and $\{C_n,~S_n\}$ following from Eqs.~\ref{eq:Bmrate} and \ref{eq:Drate}.
The explicit expressions are detailed in the companion article~\cite{LHCb-PAPER-2025-063}.
The $\{C_n,~S_n\}$ parameters are shared between the descriptions of the $B$- and $D$-decay observables,
thus a joint analysis of \besiii and \lhcb data is performed to simultaneously determine the \CP observables and strong-phase parameters.

\begin{table*}[!tb]
    \caption{Summary of tag modes and corresponding observables measured in quantum-correlated $D\Db$ decays. In the notation for the observables, $S$ denotes $\decay{D}{\KS h^{\prime+} h^{\prime-}}$ decays, $T$ and $\overline{T}$ denote the Cabbibo-favored and doubly-Cabbibo-suppressed processes in (quasi-)flavor-specific tags, respectively, \CP denotes the \CP-eigenstate tags, and $S^{\prime}$ denotes self-conjugate tags.}
    \label{tab:observables_BESIII}
    {
    \renewcommand{\arraystretch}{1.2}
    \begin{center}
        \begin{tabular}{cccccc}
            \hline
Tag category &Observables &Number of observables\\
\hline
(Quasi-)flavor-specific &$\{\Nobs_{n}(S|T),~\Nobs_{n}(S|\overline{T})\}\propto\Sigma_iw_n(\mathbf{z_i})$ &$2\times(2M+1)$\\
\CP-eigenstate &$\optbar{\mathcal{N}}_{\hspace*{-0.3em}n}(S|\CP)\propto\Sigma_i\optbar{w}_{\hspace*{-0.2em}n}(\mathbf{z_i})$ &$2\times (2M + 1)$\\
Self-conjugate &$\optbar{\mathcal{N}}_{\hspace*{-0.3em}nn'}^{}(S|S^{\prime})\propto\Sigma_i\optbar{w}_{\hspace*{-0.2em}n}(\mathbf{z_{i}})\times \optbar{w}_{\hspace*{-0.2em}n'}(\mathbf{z'_{i}})$ &$4\times (2M+1)\times(2M^{\prime}+1)$\\
\hline
        \end{tabular}
    \end{center}
    }
\end{table*}

The \besiii dataset provides the dominant constraint on the strong-phase parameters. It was collected at the $\psiprpr$ resonance in $\epem$ collisions, corresponding to an integrated luminosity of 8\invfb. 
Meanwhile, \CP-violating parameters are determined from \lhcb datasets that were collected during Run~1 and Run~2 in $pp$ collisions at center-of-mass energies ${\sqs=7,8}$ and 13\tev, corresponding to an integrated luminosity of 9\invfb. The \lhcb data also provide marginal sensitivity to the strong-phase parameters through the joint analysis~\cite{LHCb-PAPER-2025-063}.

The \besiii detector~\cite{BESIII:2009fln} records symmetric $e^+e^-$ collisions 
provided by the BEPCII storage ring~\cite{Yu:2016cof} at $\sqs = 3.773\gev$. Detection efficiencies and backgrounds are estimated by simulation samples produced with a \geant-based~\cite{Agostinelli:2002hh} software package. The simulation models the beam energy spread and initial-state radiation in $e^+e^-$ annihilations with the generator {\mbox{\textsc{kkmc}}\xspace}~\cite{Jadach:2000ir,Jadach:1999vf}. 
The \lhcb detector is a 
single-arm forward spectrometer covering the pseudorapidity range 
$2 < \eta < 5$, 
described in detail in Refs.~\cite{LHCb-DP-2008-001,LHCb-DP-2014-002}.
Simulation is required to model and correct the efficiencies in \lhcb data.
In the simulation, $pp$ collisions are generated using
\pythia~\cite{Sjostrand:2007gs,*Sjostrand:2006za} 
with a specific \lhcb configuration~\cite{LHCb-PROC-2010-056}.
Decays of unstable particles
are described by \evtgen~\cite{Lange:2001uf}, in which final-state
radiation is generated using \photos~\cite{davidson2015photos}.
The interaction of the generated particles with the detector, and its response,
are implemented using the \geant
toolkit~\cite{Allison:2006ve, Agostinelli:2002hh} as described in
Ref.~\cite{LHCb-PROC-2011-006}. 
Details on the simulation samples generated for this analysis can be found in the companion article~\cite{LHCb-PAPER-2025-063}.

In the \besiii dataset, quantum-correlated $D\Dbar$ decays are selected where one $D$ meson is reconstructed in the $\decay{D}{\KS h^{\prime+}h^{\prime-}}$ or $\decay{D}{\KL h^{\prime+}h^{\prime-}}$ final state, and the accompanying $\Db$ meson is selected in a tag mode which can be classified into three categories summarized in Table~\ref{tab:observables_BESIII}. 
Selections of the final-state particles in the doubly tagged decays are identical to those in Ref.~\cite{BESIII:2025nsp}, where the binned strong-phase parameters in the $\decay{D}{\KSL\pip\pim}$ decays are measured using the same dataset.
Following the method presented in Ref.~\cite{BESIII:2025nsp}, the selected final-state particles are combined to form the doubly tagged ${\KS h^{\prime+}h^{\prime-}}$ and ${\KL h^{\prime+}h^{\prime-}}$ candidates.
 
Possible sources of background are estimated through simulated samples of generic $D$ decays. The dominant peaking backgrounds are $\decay{D}{\pip\pim h^{\prime+}h^{\prime-}}$ for $\decay{D}{\KS h^{\prime+}h^{\prime-}}$ decays and $\decay{D}{\KS(\to \pi^0 \pi^{0})h^{\prime+}h^{\prime-}}$ for $\decay{D}{\KL h^{\prime+}h^{\prime-}}$ decays. 
 The $D$-decay observables, as listed in Table~\ref{tab:observables_BESIII}, are measured with the selected signal $D\Db$ candidates. 
To consider the background contributions in these observables, the weight functions in Eq.~\ref{eq:weightfunc} are applied to the simulated background events and the weighted background events are subtracted from the $D$-decay observables.
Efficiencies obtained from \besiii simulation are applied to correct the modification of the Dalitz-plot distributions due to reconstruction and selection effects.
As an example, $w^{\rm opt}$ in the Dalitz plots of selected signal candidates in the doubly tagged ${\KS \pip\pim}$ \vs ${\KS \Kp\Km}$ configuration are shown in Figs.~\ref{fig:opt_weights_data}(a) and (b), respectively.
The background-subtracted signal yield is $564\pm 27$, and the reconstruction efficiency is $(16.62\pm 0.03)\%$.

In the \lhcb dataset, $\decay{\KS}{\pip\pim}$ decays are reconstructed in either the \emph{long} category, where both pion tracks leave hits in the vertex detector and subsequent tracking detectors, or the \emph{downstream} category, where the decay occurs downstream of the vertex detector and tracks are detected in all other tracking detectors.
The $\KS$ candidates are combined with two tracks assigned with either the pion or kaon mass hypothesis to form $\decay{D}{\KS\pip\pim}$ and $\decay{D}{\KS\Kp\Km}$ candidates, respectively. An additional kaon or pion track is then added to reconstruct $\decay{\Bpm}{D\Kpm}$ or $\decay{\Bpm}{D\pipm}$ candidates. 
Candidate selection and signal extraction follow the same procedure as the measurement using the binned phase-space approach~\cite{LHCb-PAPER-2020-019}.

The signal extraction for $B$-candidate decays is based on a fit procedure that is detailed in the companion article~\cite{LHCb-PAPER-2025-063}.
Per-event signal weights are computed with the \sPlot technique~\cite{Pivk:2004ty}, ensuring the weighted Dalitz-plot distributions represent those of the signal. The signal candidates for $\decay{B^\pm}{D\Kpm}$ decays are shown in Figs.~\ref{fig:opt_weights_data}(c) and~(d), displaying the Dalitz distributions for $\decay{D}{\KS \pip\pim}$ and $\decay{D}{\KS \Kp\Km}$ decays, respectively.
Their signal yields are ${13\,450\pm 130}$ and ${2005\pm 50}$.

\begin{figure}[!tbp]
    \begin{center}
        \includegraphics[width=1.0\textwidth]{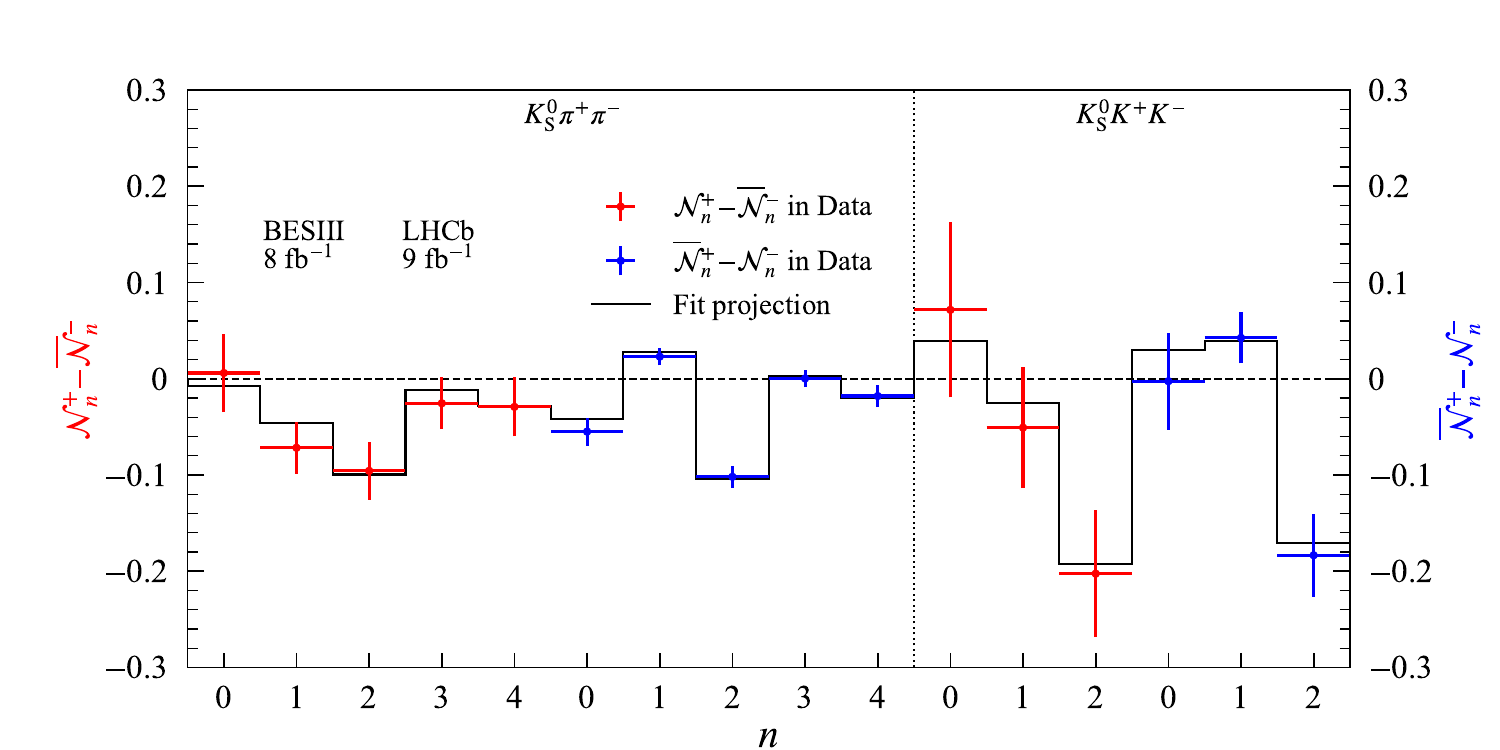}
    \end{center}
    \caption{Differences between the \CP-conjugate observables measured with the indicated $\decay{D}{\KS \pip\pim}$ and $\decay{D}{\KS \Kp\Km}$ decays.}
    \label{fig:lhcb_coef}
\end{figure}

The $B$-decay observables are measured with the selected signal events of $\decay{\Bpm}{Dh^\pm}$ according to Eq.~\ref{eq:Bobservables}. 
Per-event efficiency corrections are applied according to simulated signal decays for the $B$-decay observables. The detailed procedures of background subtraction and efficiency correction are described in the companion article~\cite{LHCb-PAPER-2025-063}. 
To reach optimal statistical precision on $\gamma$, the orders of Fourier expansions for the observables are set to be $M_{\pi} = 2$ and $M_K = 1$, as
contributions from higher-order parameters are limited under the current sample size according to pseudoexperiment studies.
Asymmetries are clearly seen in Fig.~\ref{fig:lhcb_coef} where the differences between the measured \CP-conjugate $B$-decay observables ${\Nobs_n^+-\Nobsbar_{\!n}^-}$ are plotted. 
A $\chi^2$ function, constructed using measured observables from the \besiii and \lhcb experiments, is minimized to determine the \CP-violating variables ${(\xmdk,~\ymdk)}$, ${(\xpdk,~\ypdk)}$ and ${(\xxi,~\yxi)}$ together with other nuisance parameters. 
The covariance between observables needed for the $\chi^2$ is determined from a Poisson bootstrapping method explained in Ref.~\cite{Poluektov:2017zxp}.
A weak model constraint on the difference of $\{C_n,~S_n\}$ between $\decay{D}{\KL h^{\prime+}h^{\prime-}}$ and $\decay{D}{\KS h^{\prime+} h^{\prime-}}$ decays is added to improve the determination of strong-phase parameters, which is the same approach as in the binned measurement~\cite{BESIII:2025nsp}. 
The \CP-violating parameters are determined to be
\begin{align*}
     \xmdk &= (\phantom{-}5.84\pm0.98\pm0.20) \times 10^{-2}, \\
     \ymdk &= (\phantom{-}6.74\pm1.39\pm0.25) \times 10^{-2}, \\
     \xpdk &= (   -9.68\pm1.02\pm0.26) \times 10^{-2}, \\
     \ypdk &= (   -2.42\pm1.40\pm0.31) \times 10^{-2}, \\
     \xxi  &= (   -6.57\pm2.09\pm0.30) \times 10^{-2}, \\
     \yxi  &= (   -1.12\pm2.37\pm0.20) \times 10^{-2},
\end{align*}
where the first uncertainties are statistical and the second systematic.
The minimised $\chisqndf$ corresponds to a $p$-value of 1\%, and projections of the fit results onto ${\Nobs_n^+-\Nobsbar_{\!n}^-}$ are consistent with the measurements as shown in Fig.~\ref{fig:lhcb_coef}.
Pseudoexperiments are generated to validate the fit procedure,
where the ensemble is generated with the amplitude models from Refs.~\cite{BaBar:2010nhz,BaBar:2018agf,BaBar:2018cka} and fitted with the baseline method.
Standardized-residual distributions of all \CP observables are found to be consistent with the standard normal distribution.

The \CP observables are interpreted in terms of the physics parameters ${(\gamma, \rdk, \ddk)}$ using a frequentist treatment~\cite{LHCb-PAPER-2016-032,LHCb-PAPER-2021-033}, and the resulting parameters are listed in Table~\ref{tab:parameters_lhcb}.
Considering  statistical uncertainties of the \lhcb data only, the value of $\gamma$ is determined to be $(71.3\pm 4.9){\degrees}$. 
The unbinned analysis achieves a 5\% reduction in the statistical uncertainty of $\gamma$ compared to the refitted  binned results using the same datasets and selections. Nevertheless, it does not fully recover the 15\% sensitivity loss inherent to the binned approach, primarily due to the necessary truncation of high-order Fourier expansions imposed by current statistical limitations.
Furthermore, a notable difference is observed in the central values of $\gamma$ between the binned and optimal Fourier approaches. Such a discrepancy is statistically plausible with a $p$-value of 39\%, where the correlation between two measurements is estimated from pseudoexperiments.
This indicates that the two approaches utilize different subsets of information contained within the full dataset.
The fitted central values and their uncertainties remain stable when employing higher-order Fourier expansions, indicating the robustness of the approach. 
Future analyses, with more statistics, may employ higher orders, particularly in   $M_K$, to further refine the measurement. 

\begin{table}[!tb]
    \caption{Results of the extracted parameters $(\gamma, \rdk, \ddk)$, where the uncertainties are only statistical due to the finite size of the $B$-candidate sample.}
    \label{tab:parameters_lhcb}
    \begin{center}
        \begin{tabular}{lr @{$\;$} c @{$\;$} l  r @{$\;$} c @{$\;$} l r @{$\;$} l @{$\;$} r}
            \hline
            & \multicolumn{3}{c}{$\;\gamma({\degrees})$} & \multicolumn{3}{c}{$\rdk(\times 10^{-2})$} &\multicolumn{3}{c}{$\ddk({\degrees})$}\\
            \hline
            Binned (refitted)               & $67.7$ & $\pm$& $5.1$               & $\;9.61$ & $^{+}_{-}$& $_{0.77}^{0.78}$ & $118.6$ & $^{+}_{-}$ & $^{5.2}_{5.6}$\\
            $M_\pi=2$, $M_K=1$ (baseline)   & $71.3$ & $\pm$& $4.9$               & $9.48$   & $\pm$& $0.81$                & $121.6$ & $^{+}_{-}$ & $_{5.5}^{5.3}$\\
            $M_\pi=2$, $M_K=2$              & $71.6$ & $^{+}_{-}$& $^{4.7}_{4.8}$ & $9.58$   & $^{+}_{-}$& $^{0.81}_{0.80}$ & $122.5$ & $^{+}_{-}$ & $^{5.1}_{5.4}$\\
            $M_\pi=3$, $M_K=1$              & $71.0$ & $\pm$& $4.9$               & $9.36$   & $^{+}_{-}$& $_{0.79}^{0.80}$ & $120.7$ & $^{+}_{-}$ & $_{5.6}^{5.2}$\\
            $M_\pi=3$, $M_K=2$              & $71.3$ & $^{+}_{-}$& $_{4.9}^{4.8}$ & $9.46$   & $^{+}_{-}$& $_{0.79}^{0.80}$ & $121.5$ & $^{+}_{-}$ & $^{5.1}_{5.5}$\\
            \hline
        \end{tabular}
    \end{center}
\end{table}

Systematic uncertainties on the measured observables from both experiments are found to be subleading to the statistical uncertainties. 
Following the binned measurements of the strong-phase parameters~\cite{BESIII:2025nsp} and \CP-violating parameters~\cite{LHCb-PAPER-2020-019}, the same sources of systematic uncertainty have been considered. Several additional sources are considered in this measurement due to the new analysis method.

For each source of systematic uncertainty from the analysis of \besiii data, its impact on the \CP observables is studied by performing joint fits to data or pseudoexperiments with the \lhcb component unchanged. Uncertainties from the estimation of yields of tag modes, and from background yields and distributions, are determined from pseudoexperiments.
Additional systematic uncertainties are considered, including the biases extracted from pseudoexperiments, statistical fluctuations in the flavor-tag coefficients, inputs of the hadronic parameters for flavor tags and charm-mixing parameters, the amplitude models used to determine the correlation matrices, limited statistics of the phase-space simulation samples for efficiency profiles, limited statistics of the simulation samples for signal efficiencies, and the discrepancy in resolution between data and simulation. The total systematic uncertainty from \besiii accounts for approximately 10\% of the statistical uncertainty and is subdominant compared to that from \lhcb.

Systematic uncertainties on the \CP observables from the analysis of \lhcb data derive from modifications of the fit model.
Uncertainties related to insignificant background contributions, not included in the $B$-candidate mass fits,
and physics effects, such as \CP violation in partially reconstructed $\decay{B}{D^{(*)}K^{(*)}}$ decays, are assessed using pseudoexperiments. These are generated with such effects included, then fitted with the baseline model, where the observed differences are assigned as systematic uncertainties.
Uncertainties from the fixed relative yields of partially reconstructed backgrounds, particle-identification efficiencies and mass-shape parameters are evaluated using a frequentist approach. The parameters are varied within their uncertainties, and the fits to data repeated for each variation. The resulting spread in the measured observables is taken as the systematic uncertainty. 
The effects of \CP violation in $\KS$ decay, its interactions with the detector, and $D$ mixing, are neglected in the baseline fit.
Their potential biases are quantified by generating pseudosignal samples that incorporate these effects according to Refs.~\cite{Bjorn:2019kov, Bondar:2010qs},
and fitting with the baseline model. The difference between the fitted and input values is assigned as a systematic uncertainty.
The magnitudes of the systematic uncertainties from these sources are found to be comparable to those in the binned approach~\cite{LHCb-PAPER-2020-019}.

It is assumed that there is no correlation between the mass-fit variables and Dalitz-plot distributions, as required by the \sPlot technique.
Bias from their potential correlation is assessed by investigating Dalitz-plot-dependent mass shapes from simulation.
The discrepancy between \lhcb data and simulation is corrected by comparing predicted distributions with those from $\decay{\Bpm}{D\pipm}$ data, where the background contribution is negligible.
Detector resolution is taken into account by smearing the Dalitz-plot distributions of pseudosignal according to the differences between the reconstructed and true $m^2_{\KS h^{\prime\pm}}$ in simulation.
For each source, pseudoexperiments are generated including each effect and fitted with the baseline model.
The mean biases of the \CP observables are assigned as systematic uncertainties.
The systematic uncertainty from the simulation sample size is found to be negligible.
Additionally, a systematic uncertainty is assigned to account for the small bias in the fit method, derived from a pseudoexperiment study conducted with the baseline model.
These additional sources are not dominant, hence the total systematic uncertainties remain at a level comparable to that in Ref.~\cite{LHCb-PAPER-2020-019}.

After incorporating systematic uncertainties, the parameters $(\gamma, \rdk, \ddk, \rdpi, \ddpi)$ are determined as
\begin{gather*}
    \gamma = (71.3\pm 5.0){\degrees}\,, \\
    \begin{aligned}
    \rdk &= 0.0949^{+0.0086}_{-0.0085}\,,& \rdpi &= 0.0064^{+0.0021}_{-0.0019}\,, \\
    \ddk &= (121.6^{+5.6}_{-5.9}){\degrees}\,, & \ddpi &= (311^{+17}_{-20}){\degrees}\,.
    \end{aligned}
\end{gather*}
These results are consistent with world averages~\cite{CKMfitter2005,HFLAV23,UTfit-UT} and the latest combination of \lhcb measurements~\cite{LHCb-CONF-2025-003}, in which the binned phase-space measurement is included. 
As current measurements performed with the excited \decay{\Bpm}{D^{(*)}h^{(*)\pm}} channels still rely on binned strong-phase inputs, the unbinned results contained in this Letter cannot yet be used in combination to determine a $\gamma$ average due to unclear correlations between the two approaches.
Nevertheless, this novel approach provides a complementary determination and establishes the foundation for an improved route to precision measurements of the CKM angle $\gamma$.

To summarize, the most precise single and direct measurement of the CKM angle $\gamma$ is achieved by a simultaneous analysis of the datasets of quantum-correlated $D\Db$ decays and $B$-meson decays. 
Higher-order Fourier expansions of the charm strong-phase differences could be exploited to further improve the determination of the $\gamma$ angle by taking advantage of  the 20\invfb of \psiprpr data collected by \besiii~\cite{BESIII:2024lbn} and the Run~3 dataset of \lhcb~\cite{LHCb-PII-Physics}, corresponding to an increase in integrated luminosities by a factor of about 2.5. For \lhcb, additional gains are expected from the improved trigger efficiency for hadronic channels in the Run~3 detector.
The novel approach, developed for the decay $\decay{D}{\KS\pip\pim}$, and extended to the decay $\decay{D}{\KS \Kp\Km}$, demonstrates that leveraging amplitude variations across the phase space with greater precision can enhance sensitivity in future $\gamma$ measurements. 
Moreover, the unbinned methodologies establish a blueprint for future applications to other multibody $D$ decays, \eg $\decay{D}{\Kp\Km \pip \pim}$, $\decay{D}{\pip \pim\pip \pim}$ and $\decay{D}{\pip\Km\pip \pim}$.

\section*{Acknowledgements}

\noindent We acknowledge important input from Alex Bondar, which helped to shape the analysis reported here.
The BESIII collaboration thanks the staff of BEPCII (https://cstr.cn/31109.02.BEPC) and the IHEP computing center for their strong support. This work is supported in part by National Key R\&D Program of China under Contracts Nos.\ 2023YFA1606000, 2023YFA1606704, 2022YFA1601901; National Natural Science Foundation of China (NSFC) under Contracts Nos.\ 11635010, 11935015, 11935016, 11935018, 12025502, 12035009, 12035013, 12061131003, 12192260, 12192261, 12192262, 12192263, 12192264, 12192265, 12221005, 12225509, 12235017, 12342502, 12361141819, 12375087, 12405112; the Chinese Academy of Sciences (CAS) Large-Scale Scientific Facility Program; the Strategic Priority Research Program of Chinese Academy of Sciences under Contract No.\ XDA0480600; CAS under Contract No.\ YSBR-101; 100 Talents Program of CAS; The Institute of Nuclear and Particle Physics (INPAC) and Shanghai Key Laboratory for Particle Physics and Cosmology; ERC under Contract No.\ 758462; German Research Foundation DFG under Contract No.\ FOR5327; Istituto Nazionale di Fisica Nucleare, Italy; Knut and Alice Wallenberg Foundation under Contracts Nos.\ 2021.0174, 2021.0299, 2023.0315; Ministry of Development of Turkey under Contract No.\ DPT2006K-120470; National Research Foundation of Korea under Contract No.\ NRF-2022R1A2C1092335; National Science and Technology fund of Mongolia; Polish National Science Centre under Contract No.\ 2024/53/B/ST2/00975; STFC (United Kingdom); Swedish Research Council under Contract No.\ 2019.04595; U.\ S.\ Department of Energy under Contract No.\ DE-FG02-05ER41374.

The LHCb collaboration expresses gratitude to our colleagues in the CERN
accelerator departments for the excellent performance of the LHC. We
thank the technical and administrative staff at the LHCb
institutes.
We acknowledge support from CERN and from the national agencies:
ARC (Australia);
CAPES, CNPq, FAPERJ and FINEP (Brazil); 
MOST and NSFC (China); 
CNRS/IN2P3 and CEA (France);  
BMFTR, DFG and MPG (Germany);
INFN (Italy); 
NWO (Netherlands); 
MNiSW and NCN (Poland); 
MEC/IFA (Romania); 
MICIU and AEI (Spain);
SNSF and SER (Switzerland); 
NASU (Ukraine); 
STFC (United Kingdom); 
DOE NP and NSF (USA).
We acknowledge the computing resources that are provided by ARDC (Australia), 
CBPF (Brazil),
CERN, 
IHEP and LZU (China),
IN2P3 (France), 
KIT and DESY (Germany), 
INFN (Italy), 
SURF (Netherlands),
Polish WLCG (Poland),
IFIN-HH (Romania),
PIC (Spain), CSCS (Switzerland), 
GridPP (United Kingdom),
and NSF (USA).  
We are indebted to the communities behind the multiple open-source
software packages on which we depend.
Individual groups or members have received support from
RTP (Australia), 
Key Research Program of Frontier Sciences of CAS, CAS PIFI, CAS CCEPP (China); 
Minciencias (Colombia);
EPLANET, Marie Sk\l{}odowska-Curie Actions, ERC and NextGenerationEU (European Union);
A*MIDEX, ANR, IPhU and Labex P2IO, and R\'{e}gion Auvergne-Rh\^{o}ne-Alpes (France);
Alexander-von-Humboldt Foundation (Germany);
ICSC (Italy); 
Severo Ochoa and Mar\'ia de Maeztu Units of Excellence, GVA, XuntaGal, GENCAT, InTalent-Inditex and Prog.~Atracci\'on Talento CM (Spain);
the Leverhulme Trust, the Royal Society and UKRI (United Kingdom).

\clearpage
\addcontentsline{toc}{section}{References}
\bibliographystyle{LHCb/LHCb}
\bibliography{main,LHCb/standard,LHCb/LHCb-PAPER,LHCb/LHCb-CONF,LHCb/LHCb-DP,LHCb/LHCb-TDR}
\clearpage
\centerline
{\large\bf BESIII collaboration}
\vspace{0.2cm}
\small
\noindent M.~Ablikim$^{1}$\BESIIIorcid{0000-0002-3935-619X},
M.~N.~Achasov$^{4,d}$\BESIIIorcid{0000-0002-9400-8622},
P.~Adlarson$^{83}$\BESIIIorcid{0000-0001-6280-3851},
X.~C.~Ai$^{89}$\BESIIIorcid{0000-0003-3856-2415},
C.~S.~Akondi$^{31A,31B}$\BESIIIorcid{0000-0001-6303-5217},
R.~Aliberti$^{39}$\BESIIIorcid{0000-0003-3500-4012},
A.~Amoroso$^{82A,82C}$\BESIIIorcid{0000-0002-3095-8610},
Q.~An$^{79,65,\dagger}$,
Y.~H.~An$^{89}$\BESIIIorcid{0009-0008-3419-0849},
Y.~Bai$^{63}$\BESIIIorcid{0000-0001-6593-5665}
% ...

\vspace{0.2cm} {\footnotesize\it
\noindent $^{1}$ Institute of High Energy Physics, Beijing 100049, People's Republic of China\\
% ...
$^{4}$ Budker Institute of Nuclear Physics SB RAS (BINP), Novosibirsk 630090, Russia\\
% ...
$^{31}$ INFN Sezione di Ferrara, (A)INFN Sezione di Ferrara, I-44122, Ferrara, Italy; (B)University of Ferrara, I-44122, Ferrara, Italy\\
% ...
$^{39}$ Johannes Gutenberg University of Mainz, Johann-Joachim-Becher-Weg 45, D-55099 Mainz, Germany\\
% ...
$^{63}$ Southeast University, Nanjing 211100, People's Republic of China\\
% ...
$^{65}$ State Key Laboratory of Particle Detection and Electronics, Beijing 100049, Hefei 230026, People's Republic of China\\
% ...
$^{79}$ University of Science and Technology of China, Hefei 230026, People's Republic of China\\
% ...
$^{82}$ University of Turin and INFN, (A)University of Turin, I-10125, Turin, Italy; (B)University of Eastern Piedmont, I-15121, Alessandria, Italy; (C)INFN, I-10125, Turin, Italy\\
$^{83}$ Uppsala University, Box 516, SE-75120 Uppsala, Sweden\\
% ...
$^{89}$ Zhengzhou University, Zhengzhou 450001, People's Republic of China\\

\vspace{0.2cm}
\noindent $^{\dagger}$ Deceased\\
% ...
$^{d}$ Also at the Novosibirsk State University, Novosibirsk, 630090, Russia\\
% ...

}

\centerline
{\large\bf LHCb collaboration}
\begin
{flushleft}
\small
R.~Aaij$^{38}$\lhcborcid{0000-0003-0533-1952},
A.S.W.~Abdelmotteleb$^{58}$\lhcborcid{0000-0001-7905-0542},
C.~Abellan~Beteta$^{52}$\lhcborcid{0009-0009-0869-6798},
F.~Abudin\'en$^{60}$\lhcborcid{0000-0002-6737-3528},
T.~Ackernley$^{62}$\lhcborcid{0000-0002-5951-3498},
A. A. ~Adefisoye$^{70}$\lhcborcid{0000-0003-2448-1550},
B.~Adeva$^{48}$\lhcborcid{0000-0001-9756-3712},
M.~Adinolfi$^{56}$\lhcborcid{0000-0002-1326-1264},
P.~Adlarson$^{88}$\lhcborcid{0000-0001-6280-3851},
C.~Agapopoulou$^{14}$\lhcborcid{0000-0002-2368-0147}
% ...

{\footnotesize \it

$^{1}$School of Physics and Astronomy, Monash University, Melbourne, Australia\\
% ...
$^{4}$Department of Engineering Physics, Tsinghua University, Beijing, China\\
$^{5}$Institute Of High Energy Physics (IHEP), Beijing, China\\
% ...
$^{14}$Universit{\'e} Paris-Saclay, CNRS/IN2P3, IJCLab, Orsay, France\\
% ...
$^{20}$Physikalisches Institut, Albert-Ludwigs-Universit{\"a}t Freiburg, Freiburg, Germany\\
% ...
$^{31}$INFN Sezione di Milano-Bicocca, Milano, Italy\\
% ...
$^{38}$Nikhef National Institute for Subatomic Physics, Amsterdam, Netherlands\\
$^{39}$Nikhef National Institute for Subatomic Physics and VU University Amsterdam, Amsterdam, Netherlands\\
% ...
$^{48}$Instituto Galego de F{\'\i}sica de Altas Enerx{\'\i}as (IGFAE), Universidade de Santiago de Compostela, Santiago de Compostela, Spain\\
% ...
$^{52}$Physik-Institut, Universit{\"a}t Z{\"u}rich, Z{\"u}rich, Switzerland\\
% ...
$^{56}$H.H. Wills Physics Laboratory, University of Bristol, Bristol, United Kingdom\\
% ...
$^{58}$Department of Physics, University of Warwick, Coventry, United Kingdom\\
% ...
$^{60}$School of Physics and Astronomy, University of Edinburgh, Edinburgh, United Kingdom\\
$^{61}$School of Physics and Astronomy, University of Glasgow, Glasgow, United Kingdom\\
$^{62}$Oliver Lodge Laboratory, University of Liverpool, Liverpool, United Kingdom\\
$^{63}$Imperial College London, London, United Kingdom\\
$^{64}$Department of Physics and Astronomy, University of Manchester, Manchester, United Kingdom\\
$^{65}$Department of Physics, University of Oxford, Oxford, United Kingdom\\
% ...
$^{70}$Syracuse University, Syracuse, NY, United States\\
% ...
$^{79}$Institute of Physics of  the Czech Academy of Sciences, Prague, Czech Republic, associated to $^{64}$\\
% ...
$^{82}$Faculty of Physics, Vilnius University, Vilnius, Lithuania, associated to $^{20}$\\
$^{83}$Institute of Physics and Technology, Ulan Bator, Mongolia, associated to $^{5}$\\
% ...
$^{88}$Department of Physics and Astronomy, Uppsala University, Uppsala, Sweden, associated to $^{61}$\\
$^{89}$Taras Schevchenko University of Kyiv, Faculty of Physics, Kyiv, Ukraine, associated to $^{14}$\\
% ...
$^{d}$Hangzhou Institute for Advanced Study, UCAS, Hangzhou, China\\
% ...
$ ^{\dagger}$Deceased
}
\end{flushleft}

\end{document}